# DATA COMPLIANCE IN PHARMACEUTICAL INDUSTRY
*Interoperability to align Business and IT*

Néjib Moalla, Abdelaziz Bouras, Gilles Neubert, Yacine Ouzrout
*PRISMa Laboratory Lyon 2, IUT Lumière Lyon 2, 160  Boulevard de l'Université 69500 BRON*
*{Nejib.Moalla, Abdelaziz.Bouras, Gilles.Neubert, Yacine.Ouzrout}@univ-lyon2.fr*

Abstract:	The ultimate quest in the pharmaceutical sector is *Product Quality*. We aim with this work to guarantee conformity between production and Marketing Authorizations data (Authorizations to Make to Market: AMM in Europe). These MA detail the process for manufacturing the medicine and compliance to the requirements imposed by health organisations like Food and Drug Administration (FDA) and Committee for Medicinal Products for Human use (CHMP).
The paper deals with the communication between heterogeneous information systems with different business structures and concepts. The goal is to maintain compliance of technical data in production with corresponding regulatory definition of pharmaceutical data in the MA.
Our approach to modelling present an Interoperability Framework based on a multi-layer separation to identify the organisational aspects, business trades, information systems and technologies for each involved entity into communication between these two systems. We have used RM-ODP Reference Model for the characterisation of each layer. Interoperability is guaranteed if communication of product data respects these levels.
In this work, we are also focus on the meaning of concepts and terms used by integrate business constraints in the data structure and their transformation into rules. This helps to identify integration rules to perform connections between data and their native information system and communication rules, as transformation and correspondence, to ensure the mapping through all product states.

## 1 INTRODUCTION

The pharmaceutical industry is distinguished among process industries by the need to comply with regulatory constraints imposed by organizations like Food and Drug Administration (FDA) **[1]**, Committee for Medicinal Products for Human uses (CHMP), the guidelines of International the Conference of Harmonisation (ICH) **[2]**. Further constraints are imposed by the conventions signed with national and international authorities, called Marketing Authorisation (MA) – Authorization to Make to Market (AMM) in Europe – for the manufacture of drugs.

In this operating context, the issue of product quality is one of high priority for a company in order to maintain its credibility compared to its customers.

One of the key factors of quality is the good management of product data.  Product data comes in several types and formats specific to various business trades and are supported by several heterogeneous information systems. The challenge is to enable communication among these systems and the process of guaranteeing the validity and the conformity of exchanged information. This challenge is seldom addressed systematically. Indeed, considering the complexity of information systems architectures for the production, there is a general tendency to check conformance only between the AMM files and the Standard Working Instructions (SWI).

Our Scope in this paper covers the problem of communicating product data between information system supporting the MA and the ERP for structuring production data. Deliver one product according to his description in the MA requires that we have the right information in the ERP; otherwise, we risk manufacturing a non compliant product, not delivering our product in time to respect customer commitments, destroy these product and lost money.

The pivotal problem of medical data is the absence of machine readable structures **[3]**. Most healthcare data is narrative text and often not accessible. Generally, relates works **[3][4]** have tendency to treat this problem by structuring drug information using XML standards to relating and searching drug information using topics Maps. Performing data mapping between regulatory and industrial product definition present a hard task that require regrouping effort from different sectors like regulatory affairs, industrial operations, information systems …

Some pharmaceutical industries are specialized in biologic development of medicines. The implication of a deviation in manufacturing or subcontracting can run the gamut from very minor to catastrophic.

Our challenge consists in delivering the right product data value through manufacturing states in production information system.

During manufacturing process, the product passes from one state to another. For each one, we find one or several components and we have to validate their corresponding specifications by data coming from MA information system. The following Figure (Figure 1) presents a hierarchical structuring for one product in an ERP.

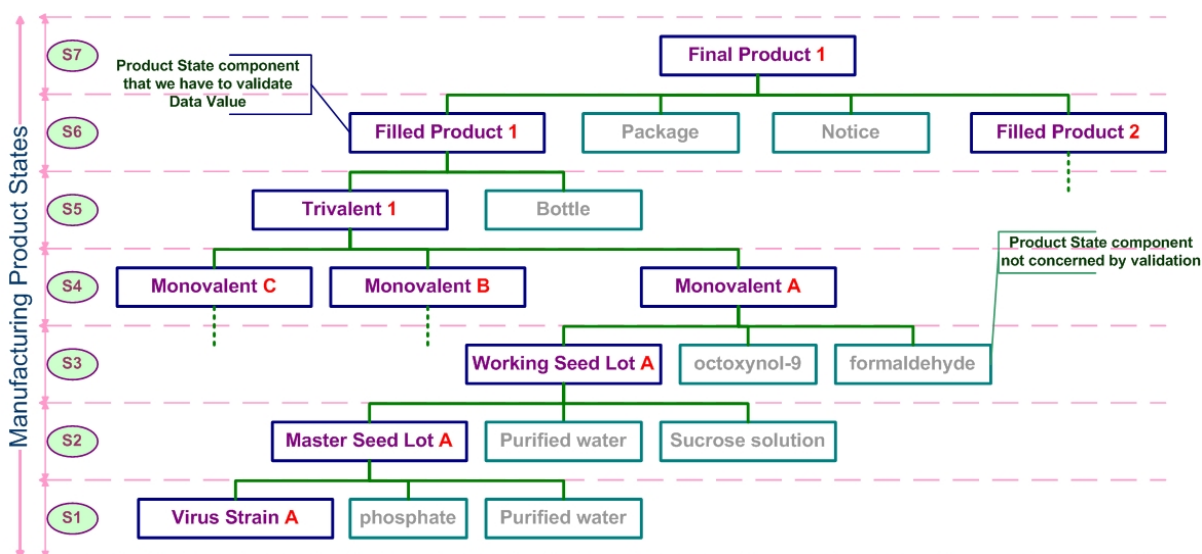

Figure 1: Manufacturing product states and state components.

When we have to ensure compliance for one data from MA to ERP, we must find and validate product data value for each component through different product states.

Our contribution in this paper is to use a modelling approach for the communication between information systems in a pharmaceutical organization. We also propose a methodology for structuring and exchanging product data while ensuring their conformance. In the following section we present some modelling approaches and adapt them to our problem. Further, we propose a data exchange structure that ensures conformity between information systems. Finally, using our approach we present a case study at Sanofi-Pasteur, a developer and producer of vaccines for human use.

## 2 MODELLING APPROACHES

### 2.1 Characteristics of pharmaceutical industry

Characteristic of the pharmaceutical industry the information for product data is compiled together from information contributed by various functional divisions that interact with each other in the creation and manufacture of the product. As we can see in figure 2, each of the following divisions contributes different types of information:

- Research division: searches for new drugs or substances that can contribute to the creation of

- new drugs. At this stage studies conducted are reported and indexed in the form reports.
- Research & Development division: conducts directed research, and is interested in the development of mixture processes of excipients, tests and stability conditions of the final solution that can be defined as a drug. The information system is used to structure data about clinical trials and tests for validity. At this stage, we begin to define an explicit product structure.
- Industrialization division: defines the industrial infrastructure which will support the production of a defined product quantity on the basis of a definition of product solution. At this stage, we start to define technical data describing the operations of product manufacture and the tools used.
- Production division: deals with planning, scheduling and follow-up of production based on the data describing industrial infrastructure and product composition. At this stage, we identify static data compared to external dynamic data like work orders and internal generated by the enterprise resources planning (ERP) like buying orders of raw material.
- Distribution division: defines the conditions for handling the product for customer delivery in accordance with the description of the conditions of manufacture delivered by R&D division. At this stage, product handling information is documented.

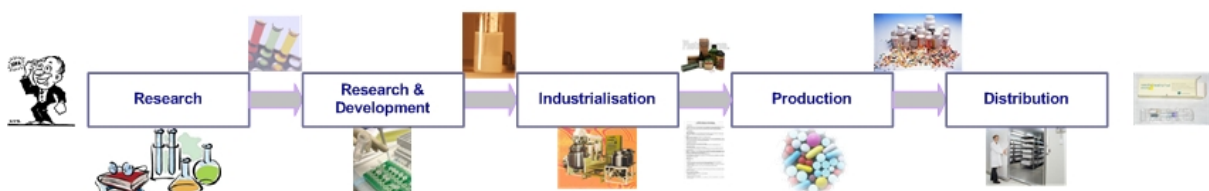

Figure 2: Information system sources in pharmaceutical industry

From one stage to another, product data are recorded using a specific structure and format. Each local information system is defined in accordance with the needs which are relevant to their business trades.

The definition of a product for pharmaceutical industry is not tied to physical shape except in the packaging stage.

The company submits to the Health Authorities entire product specifications along with documented information. These deposed documents constitute the request of Marketing Authorization. When health authorities approve this request, they give the Marketing Authorization (MA). In the delivered documents to authorities, it is necessary to present all information whose justifies product creation process, including pre-clinical tests, clinical trials, tests of validities and the appendices such as bibliography. Only after the process reaches the Industrialisation stage that MA documents can get defined.

Once approved in one country, this MA is used as reference document to manufacture product. It's considered as a contract with the authority of a given country by the company that's respects the regulatory constraints. Concerning the American market, the FDA is responsible for the checking the adequacy of the product delivered and manufacturing processes against the acquired authorization.

The major quest of each pharmaceutical company is product quality. This objective is achieved only by ensuring a better degree of conformity between existing information in these MA documents and those used for the production.

In the following sections, we will propose means to use the MA data, which can be deciphered only by pharmacists, to adapt them to logisticians needs. The approach used makes it possible to ensure interoperability between the information systems supports while satisfying some business constraints.

## 2.2 Why interoperability?

The IEEE standard computer dictionary defines interoperability as "the ability of two or more systems or components to exchange information and to use information that has been exchanged". Also, the EU Software Copyright Directive [5] defines interoperability between computing components generally to mean "the ability to exchange information and mutually to use the information which has been exchanged"[1]. This does not mean that each component must perform in the same way, or contain all of the same functionality, as every other one – interoperability is not a synonym for cloning. Rather, interoperability means that the components, which may differ in functionality, can share information and use that information to function in the manner in which they were designed to.

The European Interoperability Framework definition **[5]** identifies three separate aspects:
- Technical – is concerned with defining business goals, modelling business processes and bringing about the collaboration of administrations that wish to exchange information, but that may have a different internal organisation and structure for their operations.
- Semantic – is concerned with ensuring that the precise meaning of exchanged information is understandable by any other application not initially developed for this purpose. Semantic interoperability enables systems to combine received information with other information resources and to process it in a meaningful manner.
- Organisational – covers the technical issues of linking up computer systems and services. This includes key aspects such as open interfaces, interconnection services, data integration and middleware, data presentation and exchange, accessibility and security services

When we aim to better exchange data between information systems, we have to be sure that these interoperability types are well identified and structured. Therefore, it's necessary to identify the area of our investigation and its specifications: structures, business constraints …

## 2.3 Ways to achieve interoperability

Seeking to achieve interoperability among divisions in collaborative enterprise, we face three core challenges **[5]**:
- Heterogeneity, incoherent information perspectives, systems and software infrastructures, working practices, etc among divisions
- Need for Flexibility, due to need for change and exception handling following variation in registrations and commitments
- Complexity, the richness of interdependencies with and among divisions, their activities, resources and skills.

Heterogeneity, need for flexibility and complexity must be managed at different levels:
- Knowledge, approaches, methods and skills needed for innovation, problem solving and work performance, the shared language and frames of reference needed for communication.
- Process, the planning, coordination and management of cooperative and interdependent activities of resources.
- Infrastructure, the information formats, software tools, and interoperability approaches of the participating divisions.

---

[1] Council Directive of 14 May 1991 on the legal protection of computer programmes (91/250/EEC);

Table 1: The resulting problem space

|  | Knowledge | Process | Infrastructure |
|---|---|---|---|
| **Heterogeneity** | Communication, establishing a common languages and meaning across division and disciplines | Process diversity, negotiating different rules and procedures between the partners | Interoperability across companies' knowledge spaces and enterprise architectures (Business, Knowledge Software). |
| **Complexity** | Integrate capabilities, form effective teams across different local cultures. Align views with contents and context among and between stakeholders and people. | Work management and planning, task assignment, coordination and monitoring of activities and tasks across projects, partners and networks, dealing with uncertain interdependencies among several concurrent activities. | Enterprise architectures, managing project and systems portfolios; providing new model-driven approaches for solutions design and development; avoiding futurities (unmanageably complex systems) |
| **Flexibility** | Learning, partners must be able to improve practice based on common experience | Supporting both structured and ad-hoc work (with evolving plans); Handling unforeseen exceptions | Customised and personalised support; Rapid formation of VEs, allowing partners to join along the way |

When we try to ensure interoperability between information systems in regards of main challenges through these levels, we need to identify first different concerned entities in the company, their characteristics and link between them. To perform this task, we have to find an enterprise modelling approach bringing out all actors, processes, business constraints, work practices, etc. and identifying ties between them.

## 2.4 Enterprise Modelling Approaches

We can find in literature three main enterprise modelling approaches:

*Enterprise Framework and architectures*; define a framework as a fundamental structure which allows defining the main sets of concepts to model and build an enterprise. We describe two types of frameworks: those for integrating enterprise modelling (such as Zachman, CIMOSA, etc.) and frameworks for integrating enterprise applications (such as ISO 15745, the missing approach, etc.)

*Enterprise Modelling language*; can be define as the art of "externalising" enterprise knowledge by representing the enterprise in term of its organisation and operations. There are two categories of languages: those defined at high level of abstraction as Constructs for enterprise modelling (for example, EN/ISO 19440, ODP, UEML,GRAI,…) which are independent of the technology of implementation; and languages more related to a specific technology such as INTERNET technology based languages (example, ebXML, etc.).

*Enterprise Knowledge space*; based on a new technology called Active Knowledge Models (AKM) and implement types of views from the knowledge dimensions that contain mutual and complex dependencies of domains.

As combination of these modelling approaches, RM-ODP model (Open Distributed Processing – Reference Model) **[6] [7]** presents a methodology for the structuring of distributed services of data processing carried out in an environment of heterogeneous information systems.

Our problem is to combine these model objectives in the formalization of the systems functionalities to ensure communication independently of the tools for implementation. In the RM-ODP model, the modelling concepts, that contributes to this formalization, cover:

- Basic modelling concepts: the basic concepts are concerned with existence and activity: the expression of what exists, where it is and what it does.
- Specification concepts: addressing notions such as type and class that are necessary for reasoning about specifications, the relations between specifications, provide general tools for design, and establish requirements on specification languages.
- Structuring concepts: builds on the basic modelling concepts and the specification concepts to address recurrent structures in

distributed systems, and cover such concerns as policy, naming, behaviour, dependability and communication.

To these concepts, it is necessary to analyse the company according to several view points which influence the information structure. The view points included are:

- The enterprise viewpoint: is concerned with the purpose, scope and policies governing the activities of the specified system within the organization of which it is a part;
- The information viewpoint: is concerned with the kinds of information handled by the system and constraints on the use and interpretation of that information;
- The computational viewpoint: is concerned with the functional decomposition of the system into a set of objects that interact at interfaces – enabling system distribution;
- The engineering viewpoint: is concerned with the infrastructure required to support system distribution;
- The technology viewpoint: is concerned with the choice of technology to support system distribution.

## 2.5 How achieve interoperability can ensure compliance?

The modelling of the interconnections between information systems using these concepts, and various view points, makes it possible to ensure a structuring in accordance with definite objectives. This structuring is defined in information system urbanization (cf. project SUSIE **[8]**). Figure 3 presents these various points of view as level of abstraction.

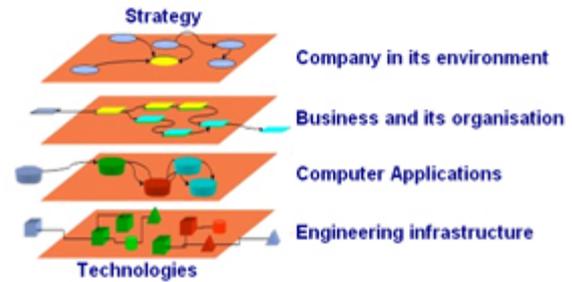

Figure 3: levels of information structuring

Interoperability between information systems could be defined more finely if this type of modelling is undertaken. In an industrial framework, structuring business knowledge in an information processing system does not imply facilitation of communication with another business system. Data interpretation changes according to the business and the challenge is in the ability to preserve information semantics when communicate across these levels. These problems also appear when we communicate product data coming from two different structuring perspectives. A proposition of communication architecture between two information systems is presented in figure 4.

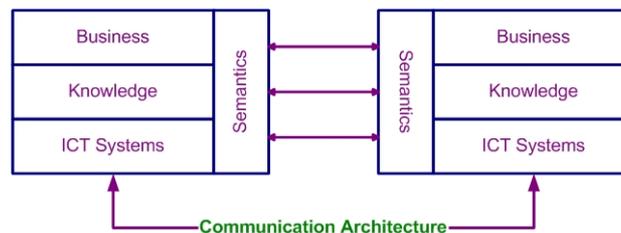

Figure 4: communication architecture between heterogeneous information systems

Building similar interoperability architecture can align Business, Knowledge and IT through semantic framework to ensure compliance when exchanging data. In the following section, we will be building our decomposition to present a deployment of an architecture of communication adapted to our context.

## 3 METHODOLOGY TO ENSURE CONFORMANCE

In our context, the objective to establish communication between information systems is to ensure the conformity of the product data in one system in relation to each other. Based on the description of information in an MA, it is necessary to return product data values to the ERP useful for the production.

## 3.1 Information system to communicate

In our scope, we identify these two entities involved in communication:
- MA information system: are generally managed by the regulatory affairs division and constitute a regrouping of information. A MA is composed of electronic documents coming from several sources and contain for example scanned documents, reports and attached papers. The semantic structuring of these authorisations provides a format and contents harmonized according to a pharmaceutical vision. It follows a specific format called Common Technical Document (CTD) **[9] [10]**, defined by the ICH. The Figure 5 presents the model of information structuring. It is a 5 modules structure. The fives modules in the figure 2 are a) specific customer information (module1), b) summary module which contains information updated from version to another (module2) c) product quality (module 3) d) non clinical study reports (module 4) and e) clinical studies reports (module 5).
The specificity of the levels of information in the MA is not significant for the retrieval and the use of the data for production division. Only pharmacists seek information from regulatory data to answer requests for product data consultation. Faced with very large number of MA documents that run into thousands of pages, makes it very difficult to answer requests for data validation.
- Production information system: was structured data necessary for the production. The ERP (Enterprise Resource Planning) system manages this data and regrouping complex functionalities of "provisioning and scheduling" and generates new dynamic data based on product data definition. So, non-in conformity product data definition, invariably leads to the manufacture of a non conforming product.

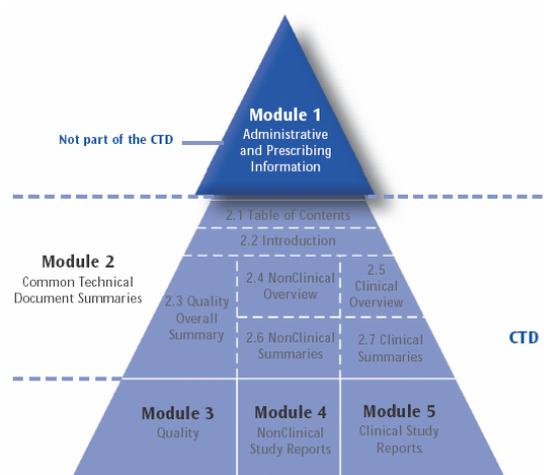

Figure 5: MA structure: CTD format **[10]**

Each of these divisions presents a specific vision of the product with local knowledge tied to business rates. It's a big challenge to communicate these two systems due to the complexity of bringing together product definition data coming from heterogeneous information systems.

To ensure this conformity at product data definition level, it is necessary to define a communication platform to include all perspectives, in particular, organisational, business, informational, and technical **[11]**. Our purpose is to present data exchange architecture allowing as to translate product data definition from regulatory systems to production one.

## 3.2 Which data we need to translate?

When we analyze information in these two systems, it's difficult to find a common data product structure. Between pharmaceutical and operational scope, we don't find necessary the same type or meaning of information. Thus, it's necessary to look first for defining what kind of information we need to ensure conformance. When fixed, we analyze product structure in each system to find issues of communication.

The CTD format of MA presents in module 3 information about product quality. In this module, we can find a pharmaceutical description of the product and its various states of manufacture. Similar in production, we define another collection of product states. These states are not necessarily coherent among them or are significant from one product to another. The best issues that we find to communicate product information from MA to ERP repose on structuring product data by states. For example, in biologic pharmacy, we define two families of product state: biologic states and pharmaceutics states, from seeds to final product states. The product structure is defined in these two systems as a specific series of product states. Our problem was translated to ensure conformance of data values for each product states from one system to another. But there are not the same definition of product states and not the same data semantic too. For example the shelf life of an intermediate substance is 3 years at storage temperature of -70°C if it was preserved as is, but -20°C if it is lyophilised. The finished product has a 1 year shelf life at 5 °C storage temperature.

We assume that the product has a fixed number of states on an information system. It's necessary to identify from production and regulatory information system, the entire specification of each state. It gathers for each stage, the value to be validated, the rules defined formally to extract data from an information system and informal business constraints helping to ensure the communication.

We call "product state reference frame", the structuring of one product datum that assign for each product state, the data value, rules applied to extract data from information system, and business constraint helping to understand the choice of data value. For each product state, we need to define also same component of the bill of materials of this state. For example, when our product present at final state two bottles, we need to specify shelf life for these two entities. For other example of this product state, this information is not significant.

The application of this reference frame to product data consists in seeking, for one data values, all states in accordance with rules and business constraints already identified. Figure 6 illustrates this structuring.

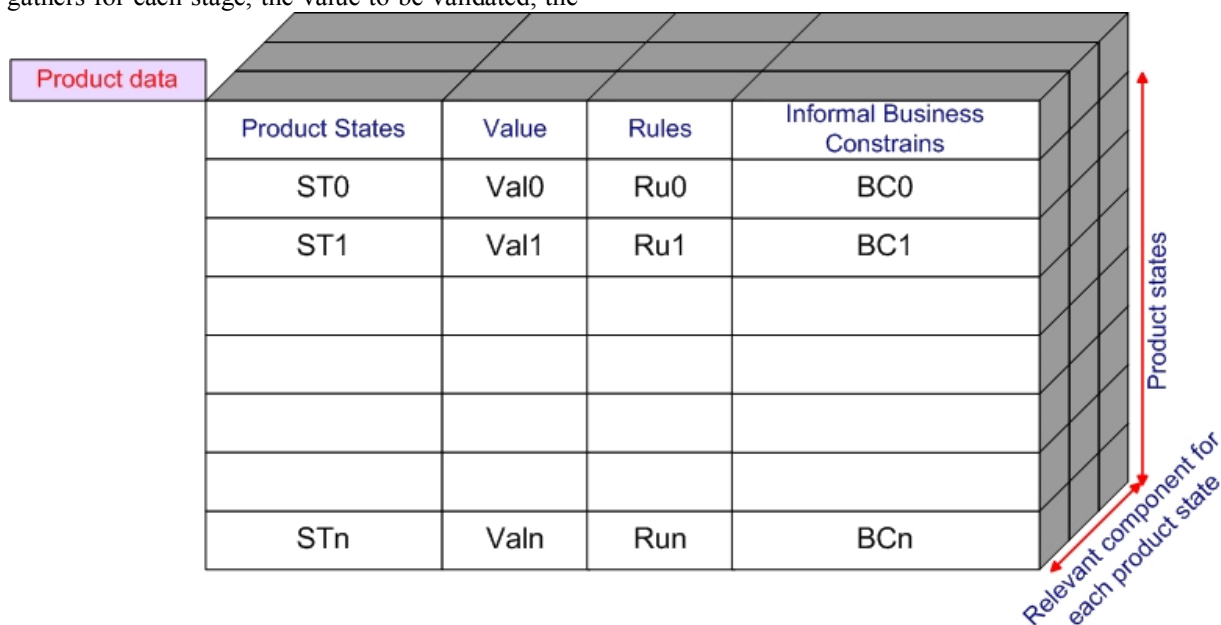

Figure 6: product states reference frame

This reference frame represents the data profile in an information system. It must be updated during a potential structural modification and can be published in the organization to ensure better comprehension and exploitation of product data.

Each line of this reference frame contains a state of product, the value to be validated, the rules which allow the extraction and the transformation of the data and business constraints for the application of these rules. For the information system of the production, we also find rules allowing for the integration of the new value.

### 3.3 Rules definition

The definition of the rules is a tedious phase and requires defining rules at three levels:

#### 3.3.1 Information system rules in production

They are rules to specify when extract or insert data into the information system. Difficulty arises when attempting to insert data. Indeed, the information systems for production are characterized by the re-use of product states information. Locking to two drugs, it is extremely probable to have the same excipients in the pharmaceutical solution. In this case, there are invariably one or more specific common production states with same coding in the system.

In production, following a request for modification of product state data, it is necessary to check if this state is also used for another product. Considering the complexity of architecture on information production system and interconnections between information inside, it is tiresome to seek products by a simple indication of an intermediate state. For example, such request can take two days to return an answer that we must sort to find required information. If we schematize the product states by a tree structure with graphs, the interconnection between graphs can be possible with any node except the first. Figure 7 shows an example of these interconnections. Each product has 6 states: S1 to S6.

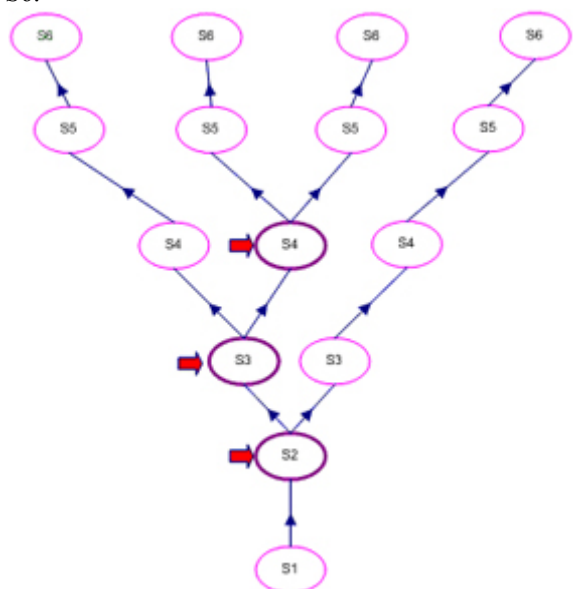

Figure 7: overlap of product states for different products in production information system

Integration rules must cover procedures of checking of the use of one state, as well as the impact of one modification if there are common states with other products. The impact of one modification or transformation on a data is sometimes governed by abstract business constraints. For example the date of manufacture of a product is calculated starting from the first valid test of stability. If we want to change the shelf life of a state, the expiry date which is equal to the date of manufacture added to the shelf life, must be revalidated. This aspect touches primarily on data transparency. This is why we planned to add informal business trade field in the reference frame for each product state to help the mapping the reference frame of the information system supporting MAs.

### 3.3.2 Mapping rules

They are rules for mapping between *product states reference frames* by establishing links between active product states. It is also a regulatory pharmaceutical responsibility that is necessary to share with production to ensure the coherence of rules. The product states are not same across information systems and certainly across reference frames. From one product to another, a state may or may not exist. We use business knowledge as a reference to create these links of communication between active states. This knowledge is indexed on both MA and production reference frames.

The mapping rules allow formalizing fields of data to be inter-connected (links **n** .. **n**) as well as transformations to use all of the values of the fields in the regulatory reference frame to generate the corresponding values in the product states reference frame of production. Figure 8 illustrates examples of connection modes. One state in the first reference frame can correspond to one or more states in the second and vice versa. To generate mapping rules, we use data and rules from the two reference frames. For example, mapping rules can be the sum, the average, the min or a data which exist only in one of the information systems.

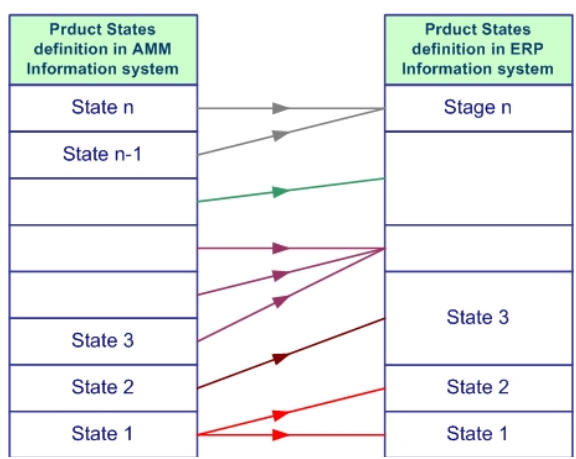

Figure 8: Mapping links

### 3.3.3 Regulatory information systems rules

According to pharmaceutical data structuring, the information system which manages the MA is not able to be directly interfaced to the regulatory product states reference frame. It is possible to have several MA for only one product, and conversely, one MA for several products. These characteristics are relocated on product states, which increase the complexity of the information retrieval. It is very frequent to find for example two product authorizations with various destinations (country) or presentations (packaging contain) and having a common product state but with different value. This difference is due to the history of the negotiations between the company and health authority about MA content.

## 4 CASE STUDY

This case study presents an illustration of a work developed with Sanofi Pasteur Company, a firm specialised on biologic development and the production of vaccines for human use. The purpose of this work is to ensure conformance, from MA to the ERP, for three data: *Site of Manufacturing, Shelf Life*, and *Storage Condition.*

All MA data was structured in e-TRAC (**E**lectronic **T**racking of **R**egistrations **a**nd **C**ommitments) information systems. Access to these data is ensured through web interface allowing us exporting defined report from *RA-Cockpit* reporting module. As presented in figure 9, we can: a) export data for one product line to create report b) distribute this report by product licence number as criteria to identify different product data c) for each product data, instantiate three reference frame for regulatory product states d) apply mapping rules to generate corresponding ERP (here SAP) product states reference frame e) use same specific criteria to data structuring in SAP to validate data comparing to those coming from SAP reference frame.

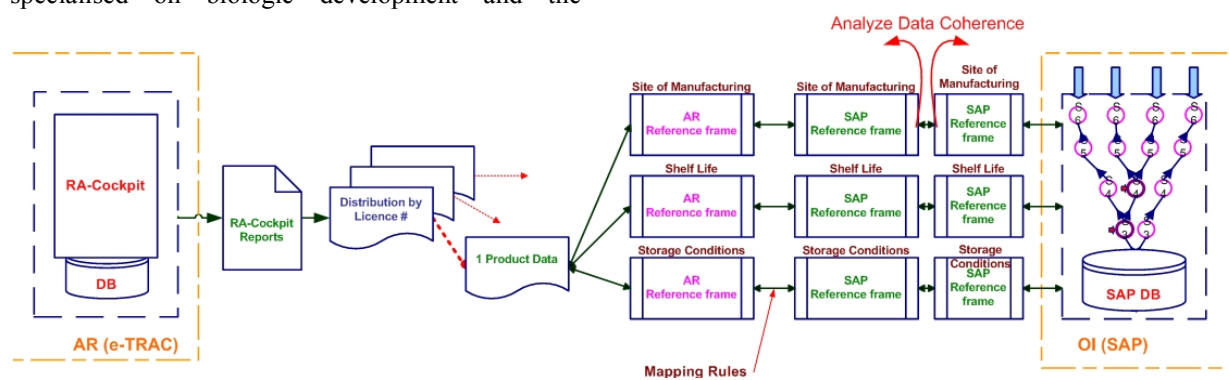

Figure 9: communication Scenario

### 4.1 Validate data in SAP

As mentioned before, it is extremely probable to have the same product state in different product decomposition, so, we can find the same value for the same product state in different SAP reference frame. In SAP systems, we identify each entity, called item, by one code. There is why, we instantiate a second SAP reference Frame with just SAP code and data value field. It's necessary to find the code of each product state. Due to specific information structuring in SAP in Sanofi-Pasteur Company, we can find the item code for the last state (final product) and use item code filiations to find code for all product states. Actually we have two SAP reference frames, one with data value coming from mapping rules with regulatory reference frame, and the second with data value and item code from SAP. We define here some rules of coherence:

- For the same product state, we have necessary the same data value, else we notify exception
- For same item codes in the second SAP reference frames we have to find the same value, else we notify exception
- It's frequent to find two or more MA or registrations that differ just by product name from one country to another. For example we can define GRIPPE vaccines in Europe but when structuring product information in regulatory information systems, we have to separate product by country. When we seek to validate our three data for grippe in Europe, we have to find the same data value in regulatory reference frame for all country in Europe, else we notify exception.

## 5 CONCLUSION

In this paper, we presented a modelling methodology for information systems especially interested in the structuring and explaining dependences between product data in the pharmaceutical field. Our objective was to ensure data compliance between two information systems, one related to the Marketing Authorizations (MA) and the other with production, through the establishment of communication architecture. Faced with information systems, we have chosen to ensure mapping between product states information along product manufacturing life cycle. In spite of differences in the business visions, the product remains at the core of information structuring in these two systems.

Applied to Grippe Line Product in Sanofi-Pasteur Company, proposed concepts provide a very interesting solution by ensuring data compliance of 94,6 % of final products for proposed data (*Site of Manufacturing, Shelf Life*, and *Storage Condition*). Some final products states in the ERP have the same definition, but not the same utility because they refer to product with different quality level. Actually we have to treat manually these specific products.

Our methodology is based in analysing information coming from MA information systems to validate him in the ERP. But what about product data in the ERP not concerned by these concepts? How to find him? And which criteria are needed to cover him by defined concepts.

We aim, with the next step of this work, to optimise our methodology of communication by adding more rules and more constraints, not only to extract or integrate data through reference frame, but between product states in one reference frame also.

## REFERENCES

**[1]** U.S Food and Drug Administration, September 2004, "Pharmaceutical CGMPS For The 21st Century — A Risk-Based Approach Final Report", Department of Health and Human Services, 32 p

**[2]** SUMMARY REPORT, 13-15 November 2003, "New Horizons and Future Challenges", *Sixth International Conference on Harmonisation, ICH 6* Osaka Japan,

**[3]** Schweigera R., Brumhardb M., Hoelzerc S., Dudecka J., 16 April 2004, "Implementing health care systems using XML Standards", *International Journal of Medical Informatics (2005) 74, 267—277*

**[4]** EBXML Technical Architecture Specification V1.0.4, ebXML Technical Architecture Project Team,16 February

**[5]** ATHENA Project (Advanced Technologies for interoperability of Heterogeneous Enterprise Networks and their Applications), February, 2005, "Second Version of State of the Art in Enterprise Modelling Techniques and Technologies to Support Enterprise Interoperability", Work package – A1.1, Version 1.3.1

**[6]** ISO/IEC 10746-3, "Information technology — Open Distributed Processing — Reference Model (RM-ODP): Architectural", First Edition, 1996

**[7]** ISO/IEC 10746-4, "Information technology — Open Distributed Processing — Reference Model (RM-ODP): Architectural semantics". First Edition 1998

**[8]** Nicolas FIGAY (EADS CCR) « PLM – de la difficulté d'intégrer et de gérer l'évolution du système d'information technique avec le progiciel » MICAD 2005, Paris

**[9]** http://www.aboutctd.com/resource.htm and http://www.ich.org/cache/html/1208-272-1.html

**[10]** Ich Harmonised Tripartite Guideline, "Organisation Of The Common Technical Document For The Registration Of Pharmaceuticals For Human Use", International Conference On Harmonisation Of Technical Requirements For Registration Of Pharmaceuticals For Human Use, November 8, 2000, 14 p

**[11]** X. Gao, A. Hayder, P. G. Maropoulos and W. M. Cheung, "Application of product data management technologies for enterprise integration", International Journal of Computer Integrated Manufacturing, Taylor & Francis, Vol 16 N° 7-8, 491 – 500, October-December 2003.